\documentclass[12pt]{article}
\usepackage{epsf}

\begin{document}

\newcommand{\gtrsim}{\mathop{}_{\textstyle \sim}^{\textstyle >}}
\newcommand{\lesssim}{\mathop{}_{\textstyle \sim}^{\textstyle <} }

\newcommand{\rem}[1]{{\bf #1}}

\renewcommand{\theequation}{\thesection.\arabic{equation}}

\renewcommand{\thefootnote}{\fnsymbol{footnote}}
\setcounter{footnote}{0}
\begin{titlepage}

\def\thefootnote{\fnsymbol{footnote}}

\begin{center}

\hfill TU-777\\
\hfill September, 2006\\

\vskip .75in

{\Large \bf 
Cosmological Constraints on Gravitino LSP Scenario
with Sneutrino NLSP
}

\vskip .75in

{\large
Toru Kanzaki$^{\rm (a)}$, Masahiro Kawasaki$^{\rm (a)}$, 
Kazunori Kohri$^{\rm (b)}$, Takeo Moroi$^{\rm (c)}$
}

\vskip 0.25in

{\em $^{(a)}$Institute for Cosmic Ray Research,
University of Tokyo, Kashiwa 277-8582, JAPAN}

\vskip 0.25in

$^{\rm (b)}${\em  Harvard-Smithsonian Center for Astrophysics, 
60 Garden Street, Cambridge, MA 02138}

\vskip 0.25in

$^{\rm (c)}${\em
Department of Physics, Tohoku University,
Sendai 980-8578, JAPAN}

\end{center}
\vskip .5in

\begin{abstract}

We study the scenario where a sneutrino is the next lightest
supersymmetric particle (NLSP) and decays into a gravitino and
standard model particles. The daughter particles such as neutrinos and
quarks cause electro- and hadronic showers and affect big-bang
nucleosynthesis.  It is found that despite a small branching ratio,
four-body processes including quarks in the final state give the most
stringent constraint on the sneutrino abundance for gravitino mass of
1 $-$ 100~GeV. Pion production by high energy neutrinos is important
when the sneutrinos decay at $\sim$ 1~sec.  We also discuss the
thermal leptogenesis in the sneutrino NLSP scenario.
  
\end{abstract}

\end{titlepage}

\renewcommand{\thepage}{\arabic{page}}
\setcounter{page}{1}
\renewcommand{\thefootnote}{\#\arabic{footnote}}
\setcounter{footnote}{0}

\section{Introduction}
\setcounter{equation}{0}
\label{sec:intro}

Among various possibilities, supersymmetry (SUSY) is a prominent
candidate for physics beyond the standard model since it possibly
solves various serious problems in particle physics and cosmology.  In
particular, from cosmology point of view, supersymmetric models
contain a good candidate for dark matter; with $R$-parity
conservation, the lightest superparticle (LSP) becomes stable and can
be dark matter.  It is widely known that, if parameters are properly
chosen, relic density of the LSP agrees with the dark matter density
suggested by WMAP \cite{Spergel:2006hy}
\begin{eqnarray}
  \Omega_{\rm CDM} h^2 = 0.105^{+0.007}_{-0.013},
  \label{Omega(WMAP)}
\end{eqnarray}
where $h$ is the Hubble constant in units of $100\ {\rm km/sec/Mpc}$.
This fact, as well as other motivations of supersymmetry, provides a
strong motivation to consider supersymmetry as a new physics beyond
the standard model.

If we consider supersymmetric models, however, several problems may
also arise.  In particular, in (local) supersymmetric models,
superpartner of the graviton, i.e., gravitino, exists.  Gravitino is a
very weakly interacting particle and it may cause  serious
cosmological problems \cite{Weinberg:zq}.  In this paper, we
concentrate on the case where the gravitino is the LSP.\footnote
{For recent study for the cases with unstable gravitino, see, for
example, \cite{Kohri:2005wn}.  (See, also,
\cite{Kohri:2001jx,Cyburt:2002uv, Jedamzik:2004er,Kawasaki:2004yh,
Kawasaki:2004qu,Ellis:2005ii}.)}
In this case, gravitino becomes a potential candidate for dark matter.

If gravitino is the LSP, the next to the lightest superparticle
(NLSP), which is assumed to be the lightest superparticle in the
minimal supersymmetric standard model (MSSM) sector, decays into
gravitino with very long lifetime.\footnote
{In this paper, we assume $R$-parity conservation.}
Importantly, the lifetime may be longer than $1\ {\rm sec}$.  Thus, in
the evolution of the universe, relic NLSPs may decay during and/or
after the big-bang nucleosynthesis (BBN).  With the decay of the NLSP,
electromagnetic and hadronic showers may be induced.  Energetic
particles produced in the showers induce photo- and hadro-dissociation
and production processes of light elements (${\rm D}$, ${\rm ^3He}$,
${\rm ^4He}$, ${\rm ^6Li}$, and ${\rm ^7Li}$).  Such processes may
significantly change the prediction of standard BBN scenario and,
consequently, resultant abundances of light elements may significantly
conflict with observations \cite{Moroi:1993mb}.  Recent detailed
analysis have shown that the model is seriously constrained when the
NLSP is neutralino or stau \cite{Feng:2004mt,Steffen:2006hw}.

However, the lightest neutralino and stau are not the only possible
candidates for the lightest superparticle in the MSSM sector.  In
particular, if the scalar neutrino becomes lightest among the
superparticles in the MSSM sector, we expect that the constraints from
BBN are drastically relaxed.  This is because the dominant decay mode
of the parent particle (i.e., sneutrino) is
$\tilde{\nu}\rightarrow\psi\nu$, and hence the daughter particles are
both weakly interacting.\footnote
{One may also consider the case where the sneutrino is the LSP and
gravitino is the next LSP.  Then, the primordial gravitino decays into
neutrino-sneutrino pair.  In this case, however, gravitino is likely
to decay also into charged particles since there should exist charged
slepton which is almost degenerate with the sneutrino.  Thus, the most
stringent constraints from BBN is expected to be from
photo-dissociation processes induced by emitted charged particles;
constraints on such cases can be read off from old studies.  (See, for
example, \cite{Jedamzik:2004er,Kawasaki:2004yh,Kawasaki:2004qu}.)
Thus, in this paper, we concentrate on the case where the gravitino is
the LSP.}
In order to discuss the BBN reaction in this class of set up, it is
crucial to understand the effects of high-energy-neutrino injection on
BBN.  Such a study was partly done in \cite{Kawasaki:1994bs}; in
\cite{Kawasaki:1994bs}, however, only photo-dissociation processes
induced by high-energy neutrino injection was discussed, and
hadro-dissociation processes were not considered.  In addition, the
sneutrino NSLP was studied in \cite{Feng:2004mt} but the processes
induced by the high energy neutrinos were ignored.

In this paper, we consider cosmological constraints on models where
(i) the scalar neutrino becomes lightest among the superparticles in
the MSSM sector, and (ii) gravitino is the LSP.  We pay particular
attention to constraints from BBN.  Organization of this paper is as
follows.  In the next section, we discuss properties of sneutrino
which plays important role in our scenario.  In Section \ref{sec:bbn},
we discuss effects of long-lived sneutrino on BBN.  Then, in Section
\ref{sec:results}, we show results of our numerical study.
Implication of our results for thermal leptogenesis scenario is
discussed in Section \ref{sec:leptogenesis}.  Section
\ref{sec:conclusion} is devoted for conclusions and discussion.

\section{Properties of Sneutrino}
\setcounter{equation}{0}
\label{sec:setup}

Before discussing cosmology, we first introduce the model we consider.
In particular, we discuss properties of sneutrino which plays important
role in our study.

Here, we adopt the particle content of the MSSM.  Important
assumptions in our study are that the lightest superparticle in the
MSSM sector is sneutrino $\tilde{\nu}$, which is in $SU(2)_L$ doublet
slepton $\tilde{L}$ and that the lightest superparticle is gravitino
$\psi$.

We first discuss when sneutrino can become the lightest superparticle
in the MSSM sector.  For this purpose, for simplicity, we adopt
$SU(5)$ grand unified model, which is strongly motivated by the fact
that, with the particle content of MSSM, three gauge coupling
constants meet with a good precision at the grand-unified-theory (GUT)
scale $M_{\rm GUT}\simeq 2\times 10^{16}\ {\rm GeV}$.  Then,
left-handed (s)leptons and right-handed down-type (s)quarks are in
$\bar{\bf 5}$ representations of $SU(5)$ while other (s)fermions are
in ${\bf 10}$ representations.

Detailed mass spectrum of superparticles depends on the mechanism of
SUSY breaking.\footnote
{In anomaly mediated SUSY breaking scenario \cite{AMSB}, gravitino
becomes much heavier than superparticles in the MSSM sector and hence
it cannot be the LSP.  Thus, in anomaly-mediated models, the mass
spectrum which we consider cannot be realized and hence our following
analysis does not have relevance in such cases.  Thus, in the
following, we will not consider anomaly-mediated models.}
In the case of gauge-mediated SUSY breaking \cite{GMSB}, mass spectrum
of MSSM superparticles are so constrained as far as the messenger
sector respects $SU(5)$.  Consequently, sneutrino does not become the
LSP in the MSSM sector.

If we consider supergravity-mediated SUSY breaking scenario, on the
contrary, soft SUSY breaking masses for sfermions arise from
(non-renormalizable) interactions in K\"ahler potential.
Consequently, in this case, there exist large number of free
parameters in the soft SUSY breaking terms.  Thus, hereafter, we
concentrate on supergravity-mediated SUSY breaking scenario.

When the soft SUSY breaking terms arise from K\"ahler interaction,
soft SUSY breaking masses for $\bar{\bf 5}$ and ${\bf 10}$ can be
independent.  In this case, all the sfermions in ${\bf 10}$
representation can become heavier than sneutrino by pushing up the
SUSY breaking masses for ${\bf 10}$.  In addition, it is usually the
case that right-handed down-type squarks becomes heavier than
left-handed slepton because of the renormalization group effects due
to gluino mass, even though their SUSY breaking masses are degenerate
at the GUT scale.  Thus, it can be easily realized that left-handed
slepton becomes the lightest among the sfermions in
supergravity-mediated SUSY breaking scenarios.  In ``conventional''
scenarios, however, it is usually the case that the bino mass becomes
smaller than the sneutrino mass.  This is due to renormalization group
effects.  Denoting the unified gaugino mass as $M_{1/2}$ and SUSY
breaking mass squared of $\bar{\bf 5}$ representation as
$\tilde{m}^2_{\bar{\bf 5}}$, electroweak-scale value of the mass
squared of $SU(2)_L$ doublet slepton is estimated as
\cite{Drees:2004jm}\footnote
{Here, we have neglected the effect of Yukawa coupling constants; it
may slightly decrease $\tilde{m}^2_{\tilde{L}}(M_{\rm weak})$.}
\begin{eqnarray}
  \tilde{m}^2_{\tilde{L}}(M_{\rm weak}) \simeq
  \tilde{m}^2_{\bar{\bf 5}}
  + 0.53 M_{1/2}^2.
\end{eqnarray}
(Here, $M_{1/2}$ and $\tilde{m}^2_{\bar{\bf 5}}$ are defined at the
GUT scale.)  In addition, bino mass is given by
\begin{eqnarray}
  M_1 (M_{\rm weak}) \simeq 0.41 M_{1/2}.
\end{eqnarray}
Thus, with the assumption of $\tilde{m}^2_{\bar{\bf 5}}\geq 0$,
$M_1^2$ becomes smaller than $\tilde{m}^2_{\tilde{L}}(M_{\rm weak})$
and sneutrino becomes heavier than bino.

However, we can easily avoid this conclusion by assuming
$\tilde{m}^2_{\bar{\bf 5}}<0$.  As we mentioned, some part of SUSY
breaking parameters originate from direct interaction between SUSY
breaking fields and observable-sector fields.  Denoting the
superfields for SUSY breaking field and observable-sector field as
$\hat{z}$ and $\hat{\phi}$, respectively, the following interaction
may exist:
\begin{eqnarray}
  {\cal L}_{\rm int} = \frac{\lambda}{M_{\rm Pl}^2} \int d^4 \theta
  \hat{z}^* \hat{z} \hat{\phi}^* \hat{\phi},
\end{eqnarray}
where $\lambda$ is a coupling constant.  After SUSY is broken, scalar
component in $\hat{\phi}$ acquires SUSY breaking mass squared as
$-\frac{\lambda}{M_{\rm Pl}^2} |F_z|^2$, where $F_z$ is the
$F$-component of $\hat{z}$.  Assuming $\lambda>0$, negative
contribution to mass squared of scalar field is obtained.
Importantly, $\lambda$ is a free parameter in K\"ahler potential and
its sign is unknown from low-energy effective theory point of view.
Thus, we assume that the $\lambda$ parameter for doublet slepton is
positive and that $\tilde{m}^2_{\tilde{L}}(M_{\rm weak})<M_1^2$.  Of
course, in this case, there exists true minimum where the doublet
slepton acquires non-vanishing vacuum expectation value.  This is not
a serious problem since the transition rate to the true minimum is
small enough so that the transition to the true minimum does not occur
in the cosmic time scale.

When the relation $\tilde{m}^2_{\tilde{L}}(M_{\rm weak})<M_1^2$ holds,
sneutrino may become the lightest superparticle in the MSSM sector in
large fraction of the parameter space.  Indeed, neglecting left-right
mixing, masses of sneutrino $\tilde{\nu}$ and left-handed charged
lepton $\tilde{l}_L$ are given by
\begin{eqnarray}
    m^2_{\tilde{\nu}} &=& 
    \tilde{m}^2_{\tilde{L}}(M_{\rm weak}) 
    + \frac{1}{2} m_Z^2 \cos 2\beta,
    \\
    m^2_{\tilde{l}_L} &=& 
    \tilde{m}^2_{\tilde{L}}(M_{\rm weak}) 
    + \left( \sin^2\theta_{\rm W} - \frac{1}{2} \right)
    m_Z^2 \cos 2\beta,
\end{eqnarray}
where $\theta_{\rm W}$ is the Weinberg angle, and $\tan\beta$ is the
ratio of vacuum expectation values of two Higgs bosons.  Thus, when
$\tan\beta>1$, $\tilde{\nu}$ becomes lighter than $\tilde{l}_L$.
Thus, hereafter, we consider the case where the sneutrino is the
lightest superparticle in the MSSM sector.  In addition, as we
mentioned, we consider the case where the LSP is the gravitino $\psi$.
In this class of set up, BBN may be affected by late-time decay of
primordial sneutrino which freezes out from the thermal bath when the
cosmic temperature drops below the sneutrino mass.

\begin{figure}[t]
    \centerline{\epsfxsize=0.5\textwidth\epsfbox{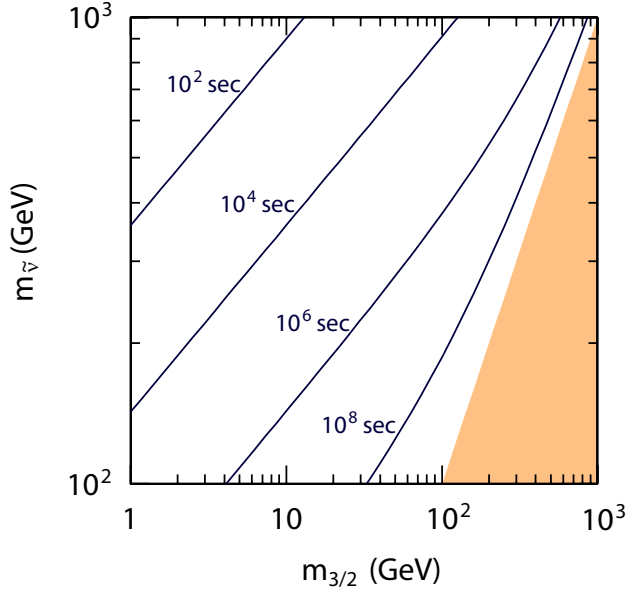}}
    \caption{Contours of constant lifetime of the sneutrino on
    $m_{3/2}$ vs.\ $m_{\tilde{\nu}}$ plane.  We have shaded the region
    with $m_{\tilde{\nu}}<m_{3/2}$, which we are not interested in.}
  \label{fig:tau_snu}
\end{figure}

Next, let us consider decay processes of $\tilde{\nu}$.  Decay of the
sneutrino is dominated by the two-body process
$\tilde{\nu}\rightarrow\psi\nu$; as we will see, decay rates for
three- and four-body decay processes are much smaller than
$\Gamma_{\tilde{\nu}\rightarrow\psi\nu}$.  Thus, the decay rate of
sneutrino is given by
\begin{eqnarray}
    \Gamma_{\tilde{\nu}} \simeq 
    \Gamma_{\tilde{\nu}\rightarrow\psi\nu} =
    \frac{m_{\tilde{\nu}}^5}{48\pi m_{3/2}^2 M_*^2}
    \left( 1 - \frac{m_{3/2}^2}{m_{\tilde{\nu}}^2} \right)^4,
    \label{Gamma(2body)}
\end{eqnarray}
where $m_{3/2}$ and $m_{\tilde{\nu}}$ are the masses of gravitino and
sneutrino, respectively, while $M_*\simeq 2.4\times 10^{18}\ {\rm
GeV}$ is the reduced Planck scale.  The lifetime of sneutrino
$\tau_{\tilde{\nu}}$ is calculated by using the two body decay rate
given in Eq.\ (\ref{Gamma(2body)}).  In Fig.\ \ref{fig:tau_snu}, we
plot contours of constant $\tau_{\tilde{\nu}}$ on $m_{3/2}$ vs.\ 
$m_{\tilde{\nu}}$ plane.  Importantly, as the gravitino mass becomes
smaller, the decay rate of $\tilde{\nu}$ becomes larger since the
interaction with the longitudinal component of the gravitino is more
enhanced.  Thus, if the gravitino mass becomes smaller than $\sim
0.1-1\ {\rm GeV}$, lifetime of sneutrino becomes shorter than $\sim 1\ 
{\rm sec}$.  In this case, primordial sneutrinos in early universe
decay before BBN starts and no constraints are obtained from BBN.  On
the contrary, with larger gravitino mass, lifetime becomes longer and
we should carefully consider hadro- and photo-dissociation processes
induced by the decay of sneutrino.

Even though the dominant decay of $\tilde{\nu}$ is via two-body
process, it is also necessary to consider three- and four-body decay
processes in the study of non-standard BBN reactions.  In particular,
processes like $\tilde{\nu}\rightarrow\psi\nu Z^{(*)}$ and
$\tilde{\nu}\rightarrow\psi lW^{(*)}$, followed by $Z^{(*)}\rightarrow
f\bar{f}$ and $W^{(*)}\rightarrow f\bar{f}'$, are important.  (Here,
the superscript ``$*$'' is for vertual particles.  In addition, $f$
represents quarks and leptons, and $l$ denotes charged lepton.)  This
is because charged and colored particles are produced only via three-
and four-body decay processes.  Those charged and colored particles
efficiently interact with background particles and induce photo- and
hadro-dissociation processes.  As we will see in the following
sections, these three- and four-body processes may significantly
change predictions of standard BBN scenario in some region of
parameter space.

\begin{figure}[t]
    \centerline{\epsfxsize=0.95\textwidth\epsfbox{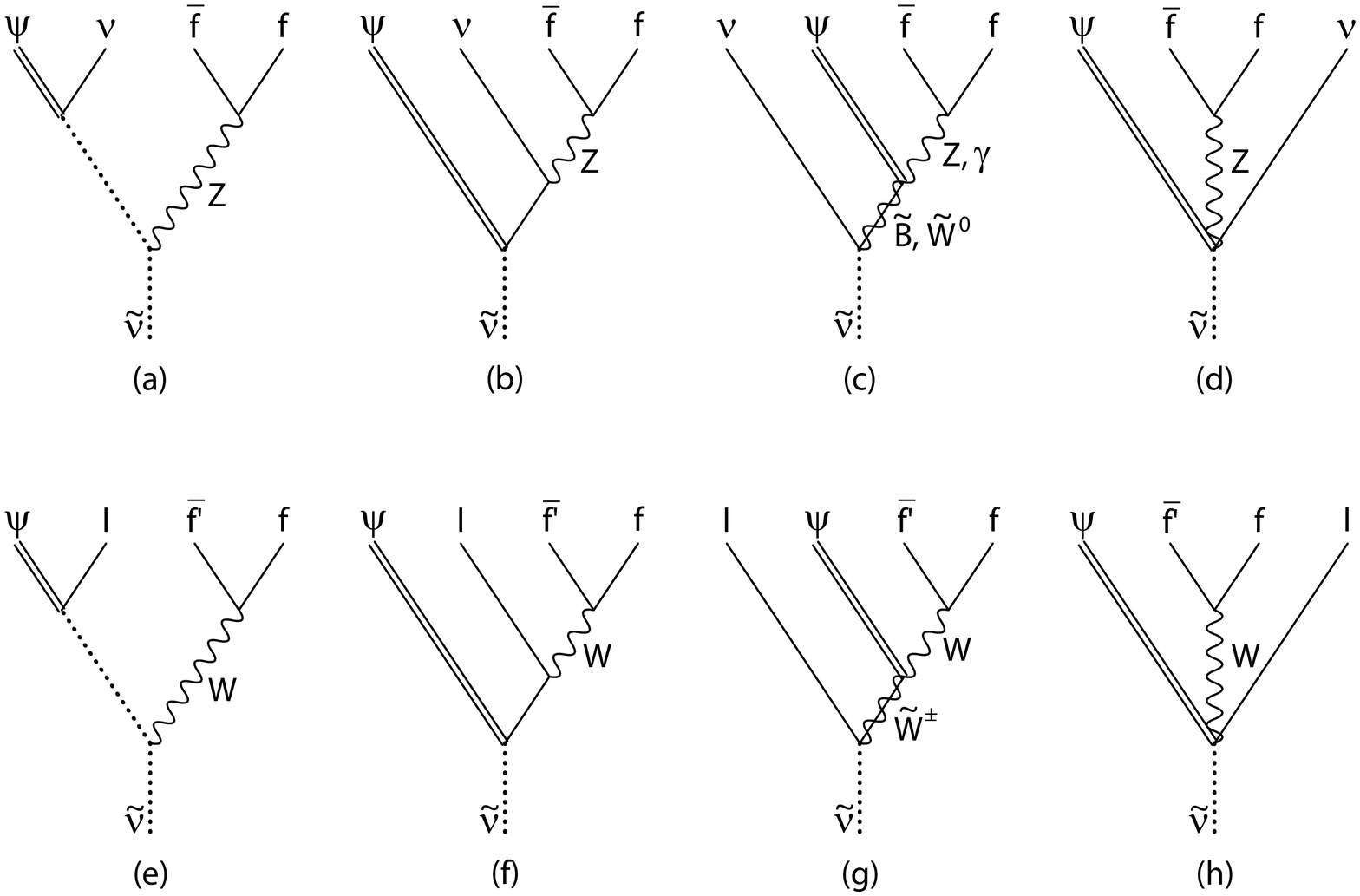}}
    \caption{Feynman diagrams for the decay processes 
    $\tilde{\nu}\rightarrow\psi\nu f\bar{f}$ ((a) $-$ (d)) and
    $\tilde{\nu}\rightarrow\psi lf\bar{f}'$ ((e) $-$ (h)).  Here,
    $\tilde{B}$, $\tilde{W}^0$, and $\tilde{W}^\pm$ represent bino,
    neutral wino, and charged wino, respectively.}
    \label{fig:feyndiag}
\end{figure}

\begin{figure}[t]
    \centerline{\epsfxsize=0.5\textwidth\epsfbox{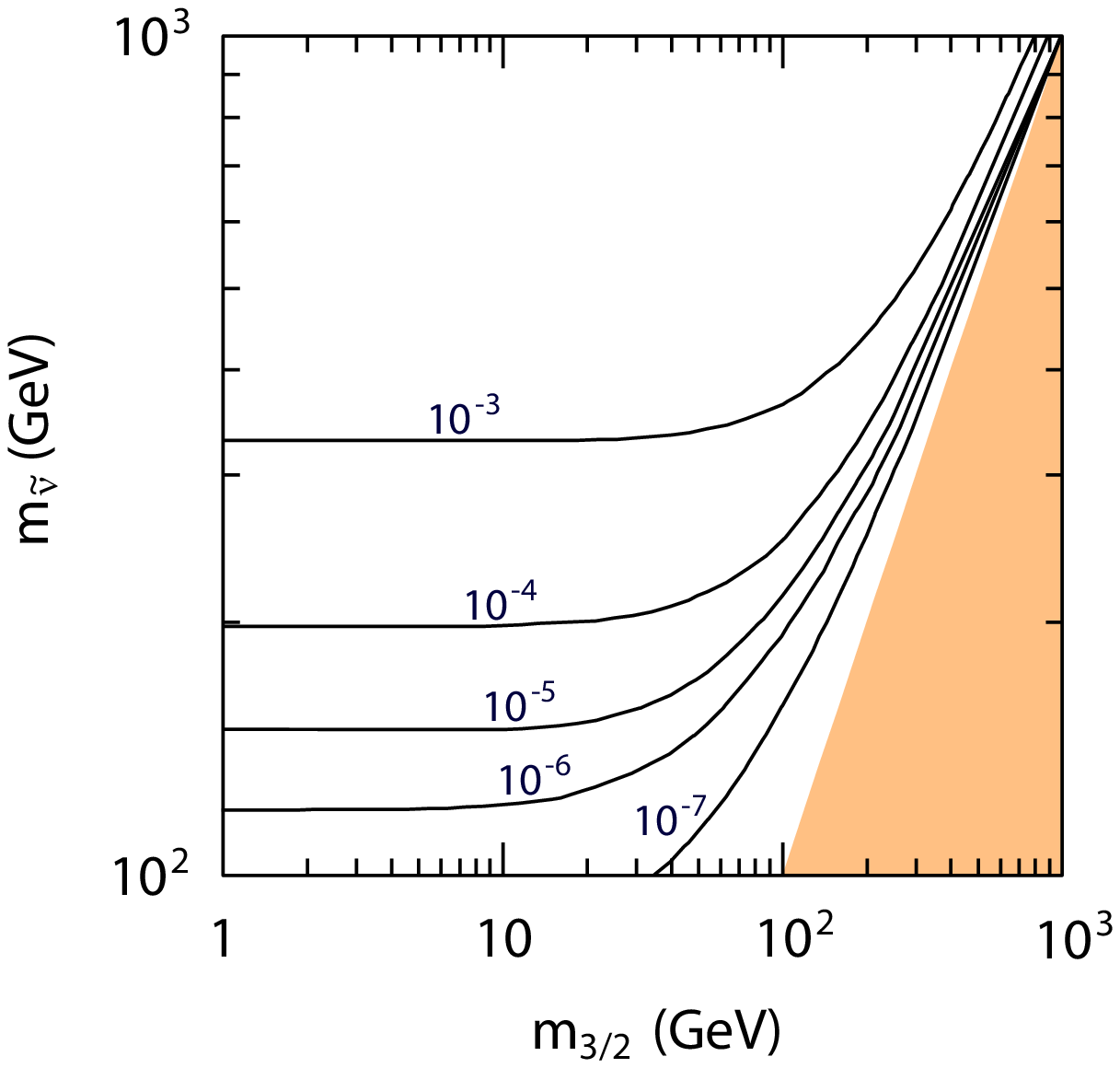}}
    \caption{Contours of constant hadronic branching ratio
    $B_h$ on $m_{3/2}$ vs.\ $m_{\tilde{\nu}}$ plane.}
    \label{fig:Bh}
\end{figure}

In order to study effects of decay modes with colored and/or charge
particles in the final state, we calculate the decays rate for the
processes $\tilde{\nu}\rightarrow\psi\nu f\bar{f}$,
$\tilde{\nu}\rightarrow\psi lf\bar{f}'$, and
$\tilde{\nu}\rightarrow\psi\nu\gamma$.  Here, we approximated that the
neutralinos and charginos are purely gaugino (or Higgsino).  In
addition, we neglect effects of Yukawa coupling constants.  Then, for
the processes $\tilde{\nu}\rightarrow\psi\nu f\bar{f}$ and
$\tilde{\nu}\rightarrow\psi lf\bar{f}'$, we take account of the
Feynman diagrams shown in Fig.\ \ref{fig:feyndiag}.  (In our analysis,
we do not use the narrow-width approximation for the productions of
``on-shell'' $Z$- and $W$-bosons; effects of ``on-shell'' $Z$ and $W$
productions are taken into account at the phase-space region where the
invariant mass of $f\bar{f}$ (or $f\bar{f}'$) system becomes close to
the gauge-boson mass.)  In Fig.\ \ref{fig:Bh}, we plot the contours of
constant hadronic branching ratio, which is defined as
\begin{eqnarray}
  B_{\rm h} = \frac{1}{\Gamma_{\tilde{\nu}}}
  \left[ \sum_q \Gamma_{\tilde{\nu}\rightarrow\psi\nu q\bar{q}}
    + \sum_q \Gamma_{\tilde{\nu}\rightarrow\psi l q\bar{q}'} \right].
\end{eqnarray}
As one can see, the hadronic branching ratio is very small.  As we
will see, however, hadronic decay processes play important role in
discussing dissociation of light elements.  We also calculate decay
rate for the process $\tilde{\nu}\rightarrow\psi\nu\gamma$, which
occurs with the diagram like Fig.\ \ref{fig:feyndiag}(c) (without
attaching $f\bar{f}$ final state).  We found, however, that the decay
rate for this process is negligibly small with the parameter set we
use in our numerical analysis.

In order to study effects of photo-dissociation processes, we also
calculate averaged ``visible energy'' emitted from the decay of one
sneutrino:
\begin{eqnarray}
    E_{\rm vis} = \frac{1}{\Gamma_{\tilde{\nu}}}
    \left[ 
        \langle 
        E_{\rm vis}^{(\tilde{\nu}\rightarrow\psi\nu\gamma)}
        \rangle
        \Gamma_{\tilde{\nu}\rightarrow\psi\nu\gamma}
        + \sum_{f} 
        \langle 
        E_{\rm vis}^{(\tilde{\nu}\rightarrow\psi\nu f\bar{f})}
        \rangle
        \Gamma_{\tilde{\nu}\rightarrow\psi\nu f\bar{f}}
        + \sum_{f} 
        \langle
        E_{\rm vis}^{(\tilde{\nu}\rightarrow\psi lf\bar{f}')}
        \rangle
        \Gamma_{\tilde{\nu}\rightarrow\psi lf\bar{f}'}
    \right],
\end{eqnarray}
where $\langle E_{\rm vis}^{(\tilde{\nu}\rightarrow\cdots)}\rangle$
are averaged energy carried away by charged particles and photon in
individual decay modes $\tilde{\nu}\rightarrow\cdots$.  We have
checked that the quantity $\langle E_{\rm
vis}^{(\tilde{\nu}\rightarrow\cdots)}\rangle$ are typically $O(10\ 
\%)$ of the sneutrino mass when $m_{3/2}\ll m_{\tilde{\nu}}$.
However, $E_{\rm vis}$ is much smaller than $m_{\tilde{\nu}}$ since
the branching ratios for events with charged particle(s) in the final
state are small.

\section{BBN with Long-Lived Sneutrino}
\setcounter{equation}{0}
\label{sec:bbn}

Now, we are at the position to discuss effects of long-lived sneutrino
on BBN.  As we mentioned, there are two types of decay modes of
sneutrino; one is two-body decay mode while the other are three-
and/or four-body ones.  Reactions caused by these decay modes are
different, so we discuss effects of these decay modes separately.

\subsection{Two-body decay}

First, we discuss effects of the dominant decay process
$\tilde{\nu}\rightarrow\psi\nu$.  The gravitino produced 
in the decay is a very weakly interacting
particle, so it is irrelevant for BBN.  

Neutrino, on the contrary, may affect abundances of light elements.
Once emitted, the energetic neutrinos may scatter off background
leptons via weak interaction.  Consequently, several kinds of
particles may be pair-produced.

First, charged leptons may be produced via the following processes:
\begin{eqnarray}
  \nu_i \bar{\nu}_{i,\rm BG} & \rightarrow &  e^- e^+ ,\\
  \nu_i \bar{\nu}_{i,\rm BG} &\rightarrow &  \mu^- \mu^+ ,\\
  \nu_{\mu} \bar{\nu}_{e,\rm BG} & \rightarrow & \mu^- e^+ ,\\
  \nu_{e} \bar{\nu}_{\mu,\rm BG} & \rightarrow & e^- \mu^+ ,
\end{eqnarray}
where $i=e,\mu,\tau$ is flavor index, and the subscript ``BG'' is for
background particles. The muons emitted in the above processes quickly
decay into electrons and neutrinos. Thus, the above processes produce
energetic electrons (and positrons) which cause electromagnetic
cascade and energetic photons in the cascade induce photo-dissociation
processes of light elements.  Effects of these processes have been
already studied in~\cite{Kawasaki:1994bs}; it was pointed out that, in
some part of the parameter space, abundances of light elements are
significantly affected if energetic neutrinos are injected.

Other possible effect is due to the production of pion pair which was
not considered in~\cite{Kawasaki:1994bs}. High energy neutrinos
scatter off the background neutrinos and electrons and produce pions
as
\begin{eqnarray}
   \nu_i \bar{\nu}_{i,\rm BG} & \rightarrow &  \pi^- \pi^+ ,\\
   \nu_i e_{i,\rm BG} & \rightarrow &  \pi^0 \pi^- . 
\end{eqnarray}
The nucleus-pion interaction rate is $\sim 10^8~\sec^{-1}\times
(T/{\rm MeV})^3$ which is larger than the decay rate of the charged
pion ($\sim 4\times 10^7~\sec^{-1}$) for $T \lesssim
1$~MeV. Therefore, the charged pions produced at $T\sim 1$~MeV scatter
off the background nuclei and change protons (neutrons) into neutrons
(protons) via
\begin{eqnarray}
     \pi^{-} p & \rightarrow & n \pi^0 ~~{\rm or}~~ n \gamma ,\\
     \pi^{+} n & \rightarrow & p \pi^0 ~~{\rm or}~~ p \gamma ,
\end{eqnarray}
which increases the $n$-$p$ ratio and synthesizes more $^4$He. On the
other hand, because of very short lifetime, the neutral pions decay
before they scatter off the background nuclei.
 
In order to estimate effects of the high energy neutrino induced
processes, we have to numerically solve the Boltzmann equation
describing the time evolution of the high energy neutrino spectrum
taking account of neutrino-neutrino (charge lepton) scattering as well
as production of charged leptons and pions.  (The details of the
Boltzmann equation and its solution will be presented elsewhere
\cite{Kanzaki}.)

\subsection{Three- and four-body processes}

Even though the branching ratio for the three- and four-body decay
processes are much smaller than $1$, such decay processes are very
important since colored and/or charged particles are directly emitted
from these decay processes.  Energetic colored and charged particles
may significantly change the prediction of standard BBN scenario.
Effects of these colored and charged particles are classified into
three categories: photo-dissociations, hadro-dissociations, and
$p\leftrightarrow n$ conversion.

Photo-dissociation processes are induced by energetic photons in
electromagnetic shower which is caused by charged particles emitted
from $\tilde{\nu}$.  With given background temperature, the
distribution function of energetic photons depends on total amount of
energy injected by particles with electromagnetic interaction, and is
insensitive to the energy spectrum of primary particles.  Thus, once
$E_{\rm vis}$ is obtained, the energy distribution of energetic
photons in the electromagnetic shower can be obtained.  Then,
photo-dissociation rates are obtained by convoluting energy
distribution function and cross sections of photo-dissociation
reactions.  For details of our treatment of photo-dissociation
processes, see \cite{KM_Photodis}.

For the study of hadro-dissociation processes, it is necessary to
obtain energy distributions of (primary) hadrons which are produced
after the hadronization of quarks emitted from $\tilde{\nu}$.  As a
first step to derive distribution functions of hadrons, we calculate
invariant-mass distribution of $q\bar{q}^{(\prime)}$ system produced
by $\tilde{\nu}\rightarrow\psi\nu q\bar{q}$ and
$\tilde{\nu}\rightarrow\psi lq\bar{q}'$.  For example, for the
neutral-current event $\tilde{\nu}\rightarrow\psi\nu q\bar{q}$, we
numerically estimate the following quantity for each quark flavor $q$:
\begin{eqnarray}
    \frac{d \Gamma_{\tilde{\nu}\rightarrow\psi\nu q\bar{q}}}
    {d m^2_{q\bar{q}}}
    = \frac{N_{\rm c}}{8 \pi^2 m_{\tilde{\nu}}}
    \int d m_{\nu q\bar{q}}^2 
    d\Phi_{\psi,(\nu q\bar{q})}
    d\Phi_{\nu,(q\bar{q})}
    d\Phi_{q,\bar{q}}
    \left| {\cal M}_{\tilde{\nu}\rightarrow\psi\nu q\bar{q}}
    \right|^2,
\end{eqnarray}
where $N_{\rm c}=3$ is the color factor, ${\cal
M}_{\tilde{\nu}\rightarrow\psi\nu q\bar{q}}$ is the matrix element,
and $m_{q\bar{q}}$ and $m_{\nu q\bar{q}}$ are invariant masses of
$q\bar{q}$ and $\nu q\bar{q}$ system, respectively.  In addition,
$d\Phi_{x,y}$ represents (infinitesimal) two-body phase space of $x$
and $y$.  For each quark flavor $q$, spectra of hadrons are calculated
by PYTHIA code \cite{Sjostrand:2006za}.  We denote the spectrum of
nucleus $N$ in the center-of-mass frame of $q\bar{q}$, which is
obtained after the hadronization of one pair of $q\bar{q}$, as
\begin{eqnarray}
    f_N^{\rm (cm)} (E_N^{\rm (cm)}; m_{q\bar{q}})
    = \left[ \frac{d N_N}{d E_N^{\rm (cm)}} \right]_{m_{q\bar{q}}},
\end{eqnarray}
where $N_N$ denotes the number of nucleus $N$.  It should be noted
that $f_N^{\rm (cm)}$ is not the spectrum in the rest frame of
$\tilde{\nu}$.  In order to take account of the effect of boost, we
numerically calculate the averaged boost factor of the $q\bar{q}$
system for as a function of $m_{q\bar{q}}$:
\begin{eqnarray}
    \bar{\gamma}_{q\bar{q}} (m_{q\bar{q}})
    = 
    \frac{1}{m_{q\bar{q}}}
    \frac
    {\int d m_{\nu q\bar{q}}^2 
    d\Phi_{\psi,(\nu q\bar{q})}
    d\Phi_{\nu,(q\bar{q})}
    d\Phi_{q,\bar{q}}E_{q\bar{q}}
    \left| {\cal M}_{\tilde{\nu}\rightarrow\psi\nu q\bar{q}}
    \right|^2}
    {\int d m_{\nu q\bar{q}}^2 
    d\Phi_{\psi,(\nu q\bar{q})}
    d\Phi_{\nu,(q\bar{q})}
    d\Phi_{q,\bar{q}}
    \left| {\cal M}_{\tilde{\nu}\rightarrow\psi\nu q\bar{q}}
    \right|^2},
\end{eqnarray}
where $E_{q\bar{q}}$ is the energy of $q\bar{q}$ system in the rest
frame of $\tilde{\nu}$.  With this quantity, we estimate the spectrum
of the nucleus $N$ in the rest frame of $\tilde{\nu}$, approximating
that the boost factor for a fixed value of $m_{q\bar{q}}$ is
universally $\bar{\gamma}_{q\bar{q}} (m_{q\bar{q}})$.  In addition,
for simplicity, we approximate that the distribution of initial $q$
(and hence $\bar{q}$) jet is isotropic in the center-of-mass frame of
$q\bar{q}$.  Then, we obtain
\begin{eqnarray}
    \left[ \frac{d N_N}{d E_N}
    \right]_{\tilde{\nu}\rightarrow\psi\nu q\bar{q}}
        = 
    \frac{1}{2\Gamma_{\tilde{\nu}\rightarrow\psi\nu q\bar{q}}} 
    \int d \cos\theta_q d m_{q\bar{q}}^2 
    \frac{\partial E_N^{\rm (cm)}}{\partial E_N}
    f_N^{\rm (cm)} (E_N^{\rm (cm)}; m_{q\bar{q}})
    \frac{d \Gamma_{\tilde{\nu}\rightarrow\psi\nu q\bar{q}}}
    {d m^2_{q\bar{q}}},
\end{eqnarray}
where, in the above formula, $E_N^{\rm (cm)}$ is related to
$E_N$ as
\begin{eqnarray}
    E_N = \bar{\gamma}_{q\bar{q}}
    \left( E_N^{\rm (cm)} + \bar{\beta}_{q\bar{q}} 
        \sqrt{E_N^{\rm (cm)2} - m_N^2} \cos\theta_q \right),
\end{eqnarray}
with $\bar{\beta}_{q\bar{q}}=\sqrt{1-\bar{\gamma}_{q\bar{q}}^{-2}}$.
Effects of charged-current events $\tilde{\nu}\rightarrow\psi l
q\bar{q}'$ are also treated in the same way.

We have calculated the spectra of $p$ and $n$ taking account of all
possible neutral- and charged-current events.  These hadrons cause
hadronic shower and induce hadro-dissociation processes.  In our
analysis, in addition, we have also estimated the number of charged
pions produced by the decay of $\tilde{\nu}$ with the same procedure
as the case of $p$ and $n$.  Such charged pions become the source of
$p\leftrightarrow n$ conversion process, which changes the number of
${\rm ^4He}$.  Once the spectra of hadrons are obtained, effects of
hadro-dissociation and $p\leftrightarrow n$ conversion are studied
following \cite{Kawasaki:2004yh,Kawasaki:2004qu}.

\section{Numerical Results}
\setcounter{equation}{0}
\label{sec:results}

Now, we are at the position to show our numerical results.  In our
analysis, we have followed the evolutions of the number densities of
light elements by numerically solving Boltzmann equations.  For this
purpose, we have modified Kawano code \cite{kawano}, including
non-standard processes discussed in the previous section:
\begin{itemize}
\item $p\leftrightarrow n$ conversion by pions produced via
    $\nu\nu_{\rm BG}\rightarrow \pi^+\pi^-$ and three- and four-body
    decays of $\tilde{\nu}$.
\item Photo-dissociation processes by charged leptons
produced via $\nu\nu_{\rm BG}\rightarrow l^+l^-$.
\item Hadro-dissociation processes by hadrons produced via
three- and four-body decay processes of $\tilde{\nu}$.
\item Photo-dissociation processes by charged particles produced via
    three- and four-body decays of $\tilde{\nu}$.
\end{itemize}
In order to derive typical constraint, in our analysis, the
primary neutrino emitted from NLSP sneutrino is treated as
equal-weight admixture of $\nu_e$, $\nu_\mu$, and $\nu_\tau$. Notice
that the most important constraints, which are from hadro-dissociation
processes, do not change even if we adopt different assumption
(although the constraints from the two-body decay of $\tilde{\nu}$ may
have slight dependence on the flavor of primary neutrino). Then, we
compare the results of our calculation with light-element abundances
inferred from observations; we calculate $\chi^2$ variable defined in
\cite{Kawasaki:2004qu} and obtained 95\ \% C.L. constraints.  The
observational constraints adopted in this paper are summarized in
Appendix \ref{app:obs}.

Effects of long-lived sneutrino on BBN depends on primordial abundance
of sneutrino.  We parameterize the primordial abundance is
parameterized by yield variable, which is defined as the ratio of
number density and total entropy density (at
$t\ll\tau_{\tilde{\nu}}$):
\begin{eqnarray}
    Y_{\tilde{\nu}} 
    \equiv \left[ \frac{n_{\tilde{\nu}}}{s} 
    \right]_{t\ll\tau_{\tilde{\nu}}}.
\end{eqnarray}
With this quantity, the number density of sneutrino is given by
$n_{\tilde{\nu}}=sY_{\tilde{\nu}}e^{-t/\tau_{\tilde{\nu}}}$.
Importantly, $Y_{\tilde{\nu}}$ depends on thermal history of the
universe, so we consider several cases.

We first consider the case where the sneutrinos are thermal relics; in
this case, we assume ``standard'' evolution of the universe when the
cosmic temperature is below the mass scale of SUSY particles $m_{\rm
SUSY}$ which is taken to be $100\ {\rm GeV}-1\ {\rm TeV}$.  When the
cosmic temperature $T$ is higher than $m_{\rm SUSY}$, all the MSSM
particles are thermalized and their number density is of order $T^3$.
As the temperature becomes lower than $m_{\rm SUSY}$, on the contrary,
density of MSSM superparticles are Boltzmann-suppressed and the number
densities of most of the superparticles become negligibly small as
$T\rightarrow 0$.  Only the exception is the number density of the
lightest superparticle in the MSSM sector, in our case, sneutrino.
With $R$-parity conservation, $\tilde{\nu}$ decays only into gravitino
and some other particle(s).    Chemical equilibrium of sneutrino is
maintained by pair annihilation process, whose rate becomes smaller
than the expansion rate of the universe at some epoch.  After this
epoch, sneutrino freezes out from the thermal bath and its number
density in the comoving volume is (almost) conserved at the cosmic
time when $t\ll\tau_{\tilde{\nu}}$.  Relic density of sneutrino in
such a case was studied in \cite{Fujii:2003nr}, and the yield variable
is given by
\begin{eqnarray}
    Y_{\tilde{\nu}} 
    \simeq 2 \times 10^{-14} \times
    \left( \frac{m_{\tilde{\nu}}}{100\ {\rm GeV}} \right).
    \label{Y_snu(thermal)}
\end{eqnarray}
Using the yield variable given above, we calculate the abundances of
light elements for fixed values of $m_{3/2}$ and $m_{\tilde{\nu}}$.
Then, comparing the results of our calculation with observational
constraints on the primordial abundances of light elements, we derive
constraints on $m_{3/2}$ vs.\ $m_{\tilde{\nu}}$ plane.  

\begin{figure}[t]
    \centerline{\epsfxsize=0.5\textwidth\epsfbox{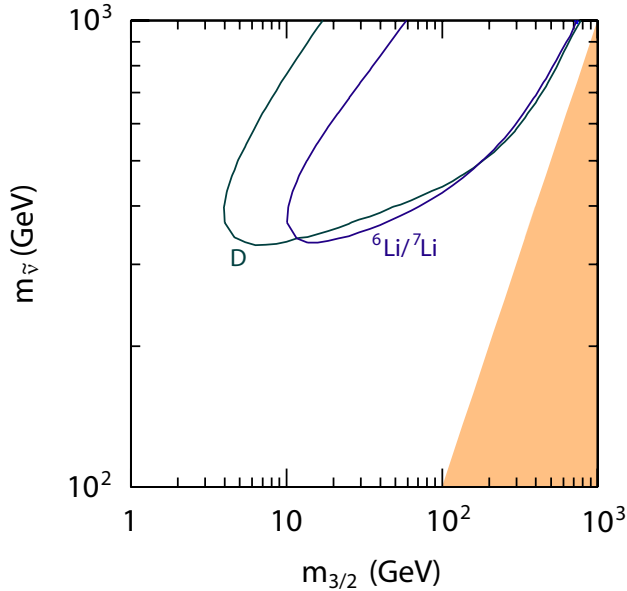}}
    \caption{Constrants on $m_{3/2}$ vs.\ $m_{\tilde{\nu}}$ plane 
    when the relic sneutrinos have thermal origin.  Regions inside the
    contours are excluded.}
  \label{fig:chi2(thermal)}
\end{figure}

In Fig.\ \ref{fig:chi2(thermal)}, we show constraints on $m_{3/2}$
vs.\ $m_{\tilde{\nu}}$ plane when the relic sneutrinos have thermal
origin.  In this case, we have found that the most important
constraints are obtained from the overproductions of ${\rm D}$ and
${\rm ^6 Li}$.  In order to understand the behavior of the
constraints, it should be noticed that the lifetime
$\tau_{\tilde{\nu}}$ becomes shorter as the gravitino mass becomes
smaller.  Consequently, when the gravitino mass is small enough,
sneutrino decays at very early stage of BBN and no constraint is
obtained.  On the contrary, when $10^2 ~{\rm
sec}\lesssim\tau_{\tilde{\nu}}\lesssim 10^7\ {\rm sec}$, the
background $^4{\rm He}$ (which we denote $\alpha_{\rm BG}$) is
effectively dissociated by the energetic hadrons produced in the
hadronic shower.  In this case, overproduction of ${\rm D}$ may occur
as a result of hadro-dissociation of $\alpha_{\rm BG}$.  In addition,
energetic ${\rm T}$ and ${\rm ^3He}$ is also produced by the
hadro-dissociation process.  Such ${\rm T}$ and ${\rm ^3He}$ become
sources of non-thermal production of ${\rm ^6Li}$ via ${\rm
T}+\alpha_{\rm BG}\rightarrow {\rm ^6Li}+n$ and ${\rm
^3He}+\alpha_{\rm BG}\rightarrow {\rm ^6Li}+p$.  As the lifetime
becomes longer, the energetic hadrons are stopped by the scattering
processes with background particles and hence the effects of
hadro-dissociations become inefficient.  In addition, we also have
found that effects of photo-dissociation are not important in this
case.  The constraint obtained here is qualitatively consistent with
the result in \cite{Feng:2004mt}, but our constraint extends to the
region with smaller gravitino mass.

So far, we have considered the case where the relic sneutrinos have
thermal origin; in such a case, the relic density of $\tilde{\nu}$ is
given by Eq.\ (\ref{Y_snu(thermal)}).  In non-standard cases, however,
relic density of $\tilde{\nu}$ may become larger.  For example, for
the case where late-decaying scalar condensation exists,
superparticles may be produced at the decay time of scalar
condensation.  If the reheating temperature after the decay of scalar
condensation is lower than $\sim 10\ {\rm GeV}$, relic density of
sneutrino becomes larger than that given in Eq.\ 
(\ref{Y_snu(thermal)}) \cite{Moroi:1999zb}.  Thus, taking
$Y_{\tilde{\nu}}$ as a free parameter, we calculate the abundances of
light elements.

\begin{figure}[t]
    \centerline{\epsfxsize=0.5\textwidth\epsfbox{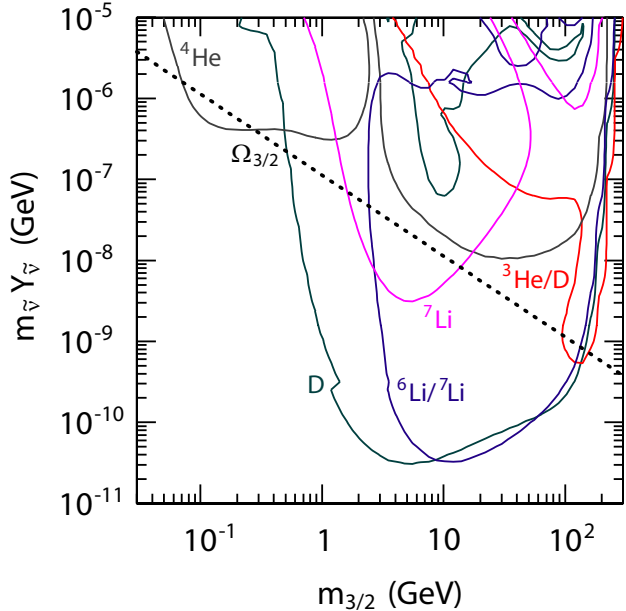}}
    \caption{Constraints on $m_{3/2}$ vs.\ 
    $m_{\tilde{\nu}}Y_{\tilde{\nu}}$ plane.  Here, we take
    $m_{\tilde{\nu}}=300\ {\rm GeV}$.  Dotted line shows the upper
    bound on $m_{\tilde{\nu}}Y_{\tilde{\nu}}$ from the overproduction
    of gravitino due to the decay of sneutrino.}
  \label{fig:chi2(nonthermal)}
\end{figure}

In Fig.\ \ref{fig:chi2(nonthermal)}, we show constraints on $m_{3/2}$
vs.\ $Y_{\tilde{\nu}}$ plane.  Here, we take $m_{\tilde{\nu}}=300\ 
{\rm GeV}$; with this value of sneutrino mass, the light-element
abundances are consistent with observational constraints if the relic
sneutrinos have thermal origin.  However, as we adopt larger value of
$Y_{\tilde{\nu}}$, some region of the parameter space is excluded.  As
one can see in Fig.\ \ref{fig:chi2(nonthermal)}, the most important
constraints are from overproduction of ${\rm ^4He}$ when
$m_{3/2}\lesssim 0.4\ {\rm GeV}$ and from overproductions of ${\rm D}$
and ${\rm ^6Li}$ when $m_{3/2}\gtrsim 0.4\ {\rm GeV}$.  In addition,
when the gravitino mass becomes close to the sneutrino mass, the
lifetime of the sneutrino becomes relatively long and
photo-dissociation of $\alpha_{\rm BG}$ becomes effective.  In this
case, overproduction of ${\rm ^3He}$ may also occur.  On the other
hand, for small gravitino mass ($\lesssim 0.4\ {\rm GeV}$) the
lifetime of the sneutrino is short ($\lesssim 40\ {\rm sec}$) and pion
production by the high energy neutrino emitted in the two-body decay
becomes significant, which leads to change $p$-$n$ ratio and ${\rm
^4He}$ overproduction.

Gravitinos are produced by the decay of sneutrinos.  Since gravitino
is stable, they contribute to the present mass density of the
universe.  Density parameter of the gravitino produced by the
sneutrino decay is obtained by
\begin{eqnarray}
    \Omega_{3/2} =  
    \frac{m_{3/2} s_{\rm now} Y_{\tilde{\nu}}}{\rho_{\rm c}},
\end{eqnarray}
where $s_{\rm now}$ is the entropy density at the present universe and
$\rho_{\rm c}$ is the critical density.  Requiring
$\Omega_{3/2}<\Omega_{\rm CDM}$, $Y_{\tilde{\nu}}$ is constrained from
above.  The bound is also shown in Fig.\
\ref{fig:chi2(nonthermal)}.\footnote
{In fact, gravitinos are also produced by scattering processes of
particles in thermal bath.  For details, see the next section.}
It should be noted that, when $m_{3/2}\gtrsim 0.1\ {\rm GeV}$, BBN
constraints on $Y_{\tilde{\nu}}$ are more stringent than that obtained
from the overproduction of the gravitino in most of the cases.

\section{Implication for Leptogenesis}
\setcounter{equation}{0}
\label{sec:leptogenesis}

In this section, we discuss implication of sneutrino-NSLP scenario for
thermal leptogenesis.  It is widely known that the present baryon
asymmetry of the universe may originate from non-equilibrium decay of
right-handed (s)neutrino which has frozen out from the thermal
\cite{Fukugita:1986hr}.

In order to realize thermal leptogenesis, however, high enough
reheating temperature is needed \cite{Buchmuller:2004nz}.  Here, we
define reheating temperature as
\begin{eqnarray}
    T_{\rm R} \equiv 
    \left( 
        \frac{10}{g_* \pi^2} M_*^2 \Gamma_{\rm inf}^2 
    \right)^{1/4},
\end{eqnarray}
where $g_*$ is the effective number of massless degrees of freedom.
(We take the MSSM value $g_*=228.75$ in our calculation.)  Then,
detailed calculation based on MSSM shows that $T_{\rm R}\gtrsim
1\times 10^{9}\ {\rm GeV}$ is required in order to generate large
enough baryon asymmetry \cite{Giudice:2003jh}.\footnote
{Reheating temperature used in \cite{Giudice:2003jh} (which we denote
$T_{\rm R}^{\rm (GNRRS)}$) is related to our definition as $T_{\rm
R}^{\rm (GNRRS)}=\sqrt{3}T_{\rm R}$.}

If the reheating temperature is as high as $\sim 10^{9}\ {\rm GeV}$,
however, large number of gravitinos are produced by scattering
processes of MSSM particles in thermal bath.  Such gravitinos usually
cause serious cosmological problems.  If gravitino is unstable,
various colored and/or charged particles are produced when gravitinos
decay, resulting in serious dissociations of light elements
during/after BBN.  If $T_{\rm R}\sim 10^{9}\ {\rm GeV}$, light-element
abundances are too much affected to be consistent with observations
unless gravitino mass is extremely large ($m_{3/2}\gtrsim 10\ {\rm
TeV}$) \cite{Kawasaki:2004qu,Kohri:2005wn}.

With stable gravitino, on the contrary, the present mass density of
gravitino may become too large \cite{Moroi:1993mb}.  Since the number
density of gravitino is approximately proportional to the reheating
temperature, we obtain upper bound on the reheating temperature.  This
fact severely constrains the scenario of thermal leptogenesis
\cite{Fujii:2003nr}.  Importantly, as the gravitino mass becomes
smaller, production of longitudinal mode of gravitino is enhanced and
upper bound on $T_{\rm R}$ becomes more stringent.

\begin{figure}[t]
  \centerline{\epsfxsize=0.5\textwidth\epsfbox{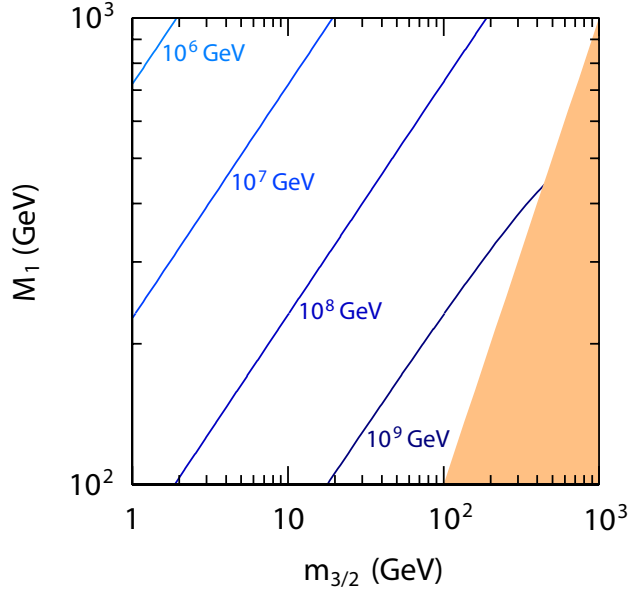}}
  \caption{Contours of constant maximal possible reheating temperature
    after inflation (or any other entropy production) in order to
    avoid overproduction of gravitino.  The horizontal axis is the
    gravitino mass while the vertical axis is bino mass.}
  \label{fig:trmax}
\end{figure}

We have numerically solved Boltzmann equations from the
inflaton-dominated epoch to radiation dominated epoch and derived
upper bound on the reheating temperature.  In our calculation, we have
assumed that, in the inflaton-dominated epoch, inflaton potential is
well approximated by parabolic potential and that inflaton is rapidly
oscillating.  Then, denoting energy densities of radiation and
inflaton as $\rho_{\rm rad}$ and $\rho_{\rm inf}$, respectively, and
the number density of gravitino as $n_{3/2}$, they obey the following
equations
\begin{eqnarray}
  \frac{d \rho_{\rm rad}}{dt} &=& -4 H \rho_{\rm rad}
  + \Gamma_{\rm inf} \rho_{\rm inf},
  \label{dot(rho_rad)}
  \\
  \frac{d \rho_{\rm inf}}{dt} &=& -3 H \rho_{\rm inf}
  - \Gamma_{\rm inf} \rho_{\rm inf},
  \\
  \frac{dn_{3/2}}{dt} &=& -3 H n_{3/2} 
  + \langle \sigma_{\rm tot} v_{\rm rel} \rangle n_{\rm rad}^2,
  \label{dot(ngra)}
\end{eqnarray}
where $H$ is the expansion rate of the universe, and $n_{\rm
rad}=\frac{\zeta (3)}{\pi^2}T^3$.  In addition, $\langle\sigma_{\rm
tot}v_{\rm rel}\rangle$ is ``thermally averaged'' total cross section
(times relative velocity), which is given in \cite{Bolz:2000fu}.  When
the gravitino mass is small, $\langle\sigma_{\rm tot}v_{\rm
rel}\rangle$ is proportional to squared of gaugino masses and,
consequently, the resultant upper bound is sensitive to the gaugino
masses.  Here, we adopt GUT relation among gaugino masses
\begin{eqnarray}
  \frac{M_3}{g_3^2} = 
  \frac{M_2}{g_2^2} = 
  \frac{3}{5} \frac{M_1}{g_1^2},
  \label{GUTrelation}
\end{eqnarray}
where $M_3$, $M_2$, $M_1$ and $g_3$, $g_2$, $g_1$ are gaugino masses
and gauge coupling constants for $SU(3)_C$, $SU(2)_L$, and $U(1)_Y$
gauge interactions, respectively.

Numerically solving Eqs.\ (\ref{dot(rho_rad)}) $-$ (\ref{dot(ngra)})
from the epoch with $H\gg\Gamma_{\rm inf}$ to the present epoch, we
calculate the density parameter of gravitino
$\Omega_{3/2}=\frac{m_{3/2}n_{3/2}}{\rho_{\rm c}}$ as a function of
reheating temperature.\footnote
{In fact, gravitinos are also produce by the decay of sneutrino.
However, we neglect such contribution since it is subdominant for most
of the cases as as we can estimate from Eq.\ (\ref{Y_snu(thermal)}).}
In deriving upper bound on $T_{\rm R}$, we require that
$\Omega_{3/2}$ be smaller than the best-fit value of the dark-matter
density reported by WMAP (see Eq.\ (\ref{Omega(WMAP)})); in other
words, the upper bound $T_{\rm R}^{\rm (max)}$ satisfies the relation
$\Omega_{3/2}(T_{\rm R}^{\rm (max)})h^2=0.105$.  Thus, for the case
where $T_{\rm R}=T_{\rm R}^{\rm (max)}$, gravitino dark matter is
realized.

The result is shown in Fig.\ \ref{fig:trmax}.  As we mentioned, the
gravitino production cross section is more enhanced with larger value
of gaugino mass and smaller value of $m_{3/2}$.  Thus, in those
limits, upper bound on $T_{\rm R}$ becomes more stringent.  Comparing
Fig.\ \ref{fig:chi2(thermal)} with Fig.\ \ref{fig:trmax}, we can see
that reheating temperature of $O(10^9\ {\rm GeV})$ is possible without
conflicting BBN constraints for $20\ {\rm GeV}\lesssim m_{3/2}\lesssim
400\ {\rm GeV}$ and $M_1\lesssim 400\ {\rm GeV}$.  (Notice that
$m_{\tilde{\nu}}$ is smaller than $M_1$ since consider the case where
sneutrino is the lightest superparticles in the MSSM sector.)  Thus in
the scenario with gravitino NLSP and gravitino LSP, thermal
leptogenesis scenario is possible in the above-mentioned parameter
region.  In this region, in addition, it should be also noticed that
gravitino is a viable candidate for dark matter.

It is important to note that the upper bound given in Fig.\ 
\ref{fig:trmax} is applicable irrespective of what the NLSP is.  Thus,
for thermal leptogenesis, gravitino mass should be $O(10\ {\rm GeV})$
or larger.  This fact gives serious constraint.  When bino or charged
slepton is the NLSP (with gravitino being the LSP), photo-dissociation
processes provide stringent constraints on the parameter region $20\ 
{\rm GeV}\lesssim m_{3/2}\lesssim 400\ {\rm GeV}$ and $M_1\lesssim
400\ {\rm GeV}$ where $T_{\rm R}\gtrsim 10^9\ {\rm GeV}$ is allowed
\cite{Feng:2004mt,Steffen:2006hw}.  In particular, photo-dissociation
processes of ${\rm ^4He}$ enhance the ratio ${\rm ^3He}/{\rm D}$ too
much.  Thus, if we adopt upper bound on the ratio ${\rm ^3He}/{\rm D}$
given in Eq.\ (\ref{r_32}), such parameter region is excluded.  Thus
we have to conclude that thermal leptogenesis is difficult for the
cases with bino NLSP and charged-slepton NLSP.  On the contrary, as we
have seen, thermal leptogenesis is possible in large parameter space
if we consider sneutrino NLSP.  Thus, the scenario with sneutrino NLSP
and gravitino LSP has an advantage in realizing thermal leptogenesis.

Before closing this section, we comment on what happens if we relax
the GUT relation on the gaugino masses.  With the GUT relation, gluino
becomes much heavier than bino.  Since the gravitino production cross
sections are proportional to squared of gaugino masses, primordial
gravitinos are mostly produced by processes with gluino in initial
and/or final states.  If the gluino mass is smaller, gravitino
production rate is suppressed for a fixed value of $M_1$.  Indeed, in
some classes unified models, GUT relation may not hold; models with
hypercolor \cite{Yanagida:1994vq} is one of such cases
\cite{Arkani-Hamed:1996jq}.  Then, with smaller gluino mass, upper
bound on the reheating temperature can become higher for a fixed value
of bino mass.  In such case, thermal leptogenesis may be possible even
with bino NLSP and stau NLSP.

\section{Conclusions}
\setcounter{equation}{0}
\label{sec:conclusion}

In this paper we have investigated the BBN constraint on the scenario
in which the sneutrino is NLSP and decays into a gravitino (i.e., LSP)
and the standard model particles. Sneutrino mainly decays into a
gravitino and a neutrino which scatters off the background neutrinos
and electrons producing pions and high energy electrons. Moreover
quarks and photons are also produced via three- or four-body decay of
the sneutrino although their branching ratios are small.  For the case
where the sneutrinos are thermal relics, the most stringent constraint
comes from overproduction of D and $^{6}$Li produced in the hadron
showers which are induced by the four-body decays. We also derived the
constraint on the sneutrino abundance assuming the sneutrinos are
produced non-thermally.  We have found that the BBN constraint is more
stringent than that from the cosmic density of the gravitino in the
wide range of gravitino mass. It is also found that pion production by
high energy neutrinos from the two-body decay becomes important when
the sneutrinos decay at $\lesssim$ 10~sec.

Since the main decay mode ($\tilde{\nu} \rightarrow  \psi\nu$)
leads to a less stringent constraint, the sneutrino NLSP scenario
allows a larger parameter space than other scenarios like 
stau NLSP. In fact, it has been shown that the thermal leptogenesis 
is realized. 

\noindent
{\it Note Added}: After finalizing the manuscript, we noticed
the paper \cite{BucCovKer} which has some overlap with our analyses.

\noindent
{\it Acknowledgments}: One of the authors (T.M.) thanks T. Asaka for
useful discussion. This work was supported in part by NASA grant
NAG5-10780 and NSF grant AST-0307433 (K.K.), and the Grant-in-Aid for
Scientific Research from the Ministry of Education, Science, Sports,
and Culture of Japan, No.\ 15540247 (T.M.).

\appendix

\section{Observational Constraints on Light Elements}
\setcounter{equation}{0}
\label{app:obs}

In this appendix, we summarize the observational constraints on
light-element abundances which we use in our analysis.  For some light
elements, we take account of recent developments in observations, so
the observational constraints adopted in this paper is different from
those used in the old study \cite{Kawasaki:2004yh,Kawasaki:2004qu}.

The primordial value of the ratio ${\rm D}/{\rm H}$ is measured in the
high red-shift QSO absorption systems.  Recently a new deuterium data
was obtained from observation of the absorption system at the red-shift
$z=2.70$ towards QSO SDSS1558-0031~\cite{O'Meara:2006mj}.  The
reported value of the deuterium abundance is $\log ({\rm D}/{\rm H}) =
-4.48\pm 0.06$. Combined with the previous
data~\cite{Tytler:1996eg,Burles:1998mk,O'Meara:2000dh,
Pettini:2001yu,Kirkman:2003uv}, it is reported that the primordial
abundance is given by
\begin{equation}
      ({\rm D}/{\rm H})_p
      = (2.82\pm 0.26) \times 10^{-5}.
\end{equation}

For the constraint on $^3$He, we adopt the same constraint as in
\cite{Kawasaki:2004yh,Kawasaki:2004qu}:
\begin{equation}
  ({\rm ^3He}/{\rm D})_p < 0.59 \pm 0.27.
  \label{r_32}
\end{equation}
Here we have used the fact that the $^3$He/D is a increasing function
of the time since D is more destroyed than $^3$He during chemical
evolution, and hence the solar abundance gives the upper limit.

The ${\rm ^4He}$ is observed in metal poor extragalactic HII regions
and the primordial abundance $Y_p$ is obtained by extrapolation of the
observed abundances to zero metallicity. For a long time relatively
low abundance had been believed for ${\rm ^4He}$. Field and
Olive~\cite{Fields:1998gv} derived $Y_p = 0.238 \pm (0.002)_{\rm stat}
\pm (0.005)_{\rm syst}$ and Izotov and Thuan~\cite{Izotov:2003xn}
obtained $Y_p=0.242 \pm 0.002$.  However, recently, Olive and
Skillman~\cite{Olive:2004kq} reanalyzed the Izotov-Thuan data and
derived higher ${\rm ^4He}$ abundance with much larger uncertainty,
$Y_p = 0.249\pm 0.09$. The similar value is obtained in more recent
analysis~\cite{Fukugita:2006xy} which we adopt here,
\begin{equation}
   Y_p = 0.250 \pm 0.004.
\end{equation}
The smaller uncertainty than that in~\cite{Olive:2004kq} is due to
different determination of the electron temperature $T_e$ 
for HII regions. The OIII lines are used in~\cite{Fukugita:2006xy} 
while \cite{Olive:2004kq} uses HeI recombination lines, 
which generally leads to large errors in $T_e$. 

Warmest metal-poor (pop.II) halo stars have almost constant ${\rm
^7Li}$ abundances (Spite plateau) independent of metallicity.  This
constant value is considered as the primordial abundance of ${\rm
^7Li}$. Bonifacio et al.~\cite{Bonifacio:2002yx} obtained
$\log_{10}({\rm ^7Li}/{\rm H})_p = -9.66 \pm 0.056$ which is
consistent with more recent value $\log_{10}({\rm ^7Li}/{\rm H})_p = -
9.63 \pm 0.06$ by Mel\'endez and Ram\'irez~\cite{Melendez:2004ni}. On
the other hand, significantly small abundance $\log_{10}({\rm
^7Li}/{\rm H})_p = -9.91 \pm 0.10$ was derived in \cite{RBOFN} where
it was claimed that there is a correlation between ${\rm ^7Li}$ and Fe
abundances due to ${\rm ^7Li}$ production by cosmic ray interactions
and one should obtain the primordial value by taking zero Fe limit.
The Fe dependence of ${\rm ^7Li}$ was also observed by Asplund et
al~\cite{Asplund:2005yt} who reported $\log_{10}({\rm ^7Li}/{\rm H})_p
= -9.90 \pm 0.06$. However, no correlation was observed in
\cite{Bonifacio:2002yx,Melendez:2004ni}. Thus, the ${\rm ^7Li}$-Fe
correlation is still an open question. Furthermore, the observed ${\rm
^7Li}$ abundance may be smaller than the primordial value if ${\rm
^7Li}$ is depleted in stars. For example, \cite{Pinsonneault:2001ub}
shows that rotational mixing leads to depletion factor $D_7$ at most
$0.3$~dex. Therefore, at present it is difficult for us to reach some
consensus.  Since the upperbound of ${\rm ^7Li}$ is important in
deriving a constraint on the sneutrino decay, we conservatively adopt
the higher value~\cite{Melendez:2004ni} and add systematic error of
0.3~dex taking depletion into account,
\begin{equation}
    \log_{10}({\rm ^7Li}/{\rm H})_p = -9.63 \pm 0.06 \pm 0.3.
\end{equation}
Here we have added $-0.3$ for systematic error because the high ${\rm
^7Li}$ abundances in \cite{Bonifacio:2002yx,Melendez:2004ni} are
different from those in \cite{RBOFN,Asplund:2005yt} by about factor of
2 ($ \simeq 0.3$~dex).

As for $^6$Li recently Asplund et al.~\cite{Asplund:2005yt} detected
$^6$Li/$^7$Li ratios in 9 metal poor halo dwarfs. In particular,
$^6$Li abundance detected in very metal poor star LP 815-43 
with [Fe/H] = -2.74 was is $^6$Li/$^7$Li $= 0.046\pm 0.022$. Such 
high $^6$Li abundance seems difficult to explain by the Galactic 
cosmic ray spallation and $\alpha$-fusion reactions and might be 
primordial. In this paper we regard it as upper bound on 
the primordial $^6$Li abundance. Moreover, the depletion in stars 
is more important for $^6$Li since $^6$Li is more fragile than $^7$Li.
The depletion factor $D_6$ for $^6$Li is related to $D_7$ as
$D_6 \simeq 2.5 D_7$~\cite{Pinsonneault:1998nf}, which leads to
$\log_{10}(^6{\rm Li}/^7{\rm Li})_p = 
1.5D_7+\log_{10}(^6{\rm Li}/^7{\rm Li})_{\rm obs}$
Taking  account of the depletion, we adopt 
\begin{equation}
   (^6{\rm Li}/^7{\rm Li})_p < 0.046\pm 0.022 + 0.084.
\end{equation}

\end{document}